\documentclass[conference]{IEEEtran}
\IEEEoverridecommandlockouts
\usepackage{cite}
\usepackage{amsmath,amssymb,amsfonts}
\usepackage{listings}
\usepackage{algorithmic}
\usepackage{graphicx}
\usepackage{textcomp}
\usepackage{xcolor}
\usepackage{url}
\usepackage{verbatim}
\usepackage{hyperref}
\usepackage{caption}
\usepackage{tabularx}
\def\BibTeX{{\rm B\kern-.05em{\sc i\kern-.025em b}\kern-.08em
    T\kern-.1667em\lower.7ex\hbox{E}\kern-.125emX}}
    
    \definecolor{lightGray}{gray}{0.9}
\begin{document}

\newcommand{\nb}[2]{
    \fbox{\bfseries\sffamily\scriptsize#1}
    {\sf\small\textcolor{red}{\textit{#2}}}
}
\newcommand\ag[1]{\nb{AG}{#1}}
\newcommand\ts[1]{\nb{ST}{#1}}


\title{OCTA-Based Biomarker Characterization in nAMD}

\author{
    \IEEEauthorblockN{
        1\textsuperscript{st} Maria Simona Tivadar\IEEEauthorrefmark{2}, 
        2\textsuperscript{nd} Ioana Damian\IEEEauthorrefmark{1}, 
        3\textsuperscript{rd} Adrian Groza\IEEEauthorrefmark{1}, 
        4\textsuperscript{th} Simona Delia Nicoara\IEEEauthorrefmark{2}
    }
    \IEEEauthorblockA{\IEEEauthorrefmark{1} Department of Computer Science, \\
        Technical University of Cluj-Napoca, 400114 Cluj-Napoca, Romania  \\ 
        European University of Technology, European Union \\ 
        \{Simona.Tivadar@student.utcluj.ro, adrian.groza@cs.utcluj.ro\}
    }
    \IEEEauthorblockA{\IEEEauthorrefmark{2} Department of Ophthalmology, \\
        “Iuliu Hatieganu” University of Medicine and Pharmacy, \\
        Emergency County Hospital, 400347 Cluj-Napoca, Romania \\ 
        \{ioana.damian@umfcluj.ro\}
    }
}

\maketitle

\begin{abstract}
We aim to enhance ophthalmologists' decision-making when diagnosing the Neovascular Age-Related Macular Degeneration (nAMD). 
We developed three tools to analyze Optical Coherence Tomography Angiography images: (1) extracting biomarkers such as mCNV area and vessel density using image processing; (2) generating a 3D visualization of the neovascularization for a better view of the affected regions; and (3) applying an ensemble of three white box machine learning algorithms (decision tree, support vector machines and DL-Learner) for nAMD diagnosis. The learned expressions reached 100\% accuracy for the training data and 68\% accuracy in testing. The main advantage is that all the learned models white-box, which ensures explainability and transparency, allowing clinicians to better understand the decision-making process.
\end{abstract}

\begin{IEEEkeywords}
Neovascular Age Related Nacular Degeneration (nAMD), Optical Coherence Tomography Angiography (OCTA), 3D visualisation, 
white-box machine learning
\end{IEEEkeywords}


\section{Motivation}
Neovascular age-related macular degeneration (nAMD)~\cite{liu2015automated} is a leading cause of blindness in people aged 50 years or older. It is characterized by the presence of choroidal neovascularization (CNV), which involves the growth of new blood vessels originating from the choroid through a break in the Bruch's membrane, extending into the sub-retinal pigment epithelium (sub-RPE) or subretinal space.

OCTA is a non-invasive imaging technique that provides a detailed view of the blood vessels in the retina and choroid, the eye's vascular layers~\cite{de2015review}. This advanced method enhances the capabilities of traditional Optical Coherence Tomography (OCT)~\cite{diag2023} by offering not only structural information but also functional insights into blood flow within the eye's intricate vascular networks, without the need for dye injections as used in conventional fluorescein angiography.

Although OCTA imaging has advantages over conventional OCT, it still produces a significant number of artifacts and vessel structures that can confuse physicians when interpreting the images. Diagnosing and evaluating the size of blood vessels, their progression over time, and their response to treatment can be challenging. Since the blood vessels at the back of the eye are so small, they often grow or shrink by only a few microns over a period of three months, making it extremely difficult for doctors to accurately assess these changes by simply examining the OCTA images.

The diagnosis of nAMD is clinically established through the identification of specific signs during a fundus examination, such as subretinal fluid, hard exudates, hemorrhages, and detachment of the retinal pigment epithelium (RPE). This diagnosis is then confirmed using fluorescein angiography, which highlights the presence, location, and size of the neovascular membrane, and through OCT, which confirms the presence of subretinal or sub-RPE fluid. 
In contrast, OCTA is the only imaging technique that allows direct visualization of the choroidal neovascular membrane, enabling qualitative analysis of its shape and pattern. However, a quantitative approach that involves measuring multiple parameters remains challenging due to technical limitations.
 
\begin{table}
    \centering
    \caption{Imaging techniques senses various biomarkers}
    \label{tab:img}
    \begin{tabular}{ll}
        Imaging technique & Targeted biomarkers \\ \hline
        Color fundus &  hemorrhages \\
        OCT & elevation of the RPE, subretinal fluid, hard exudates \\
        FA & speckled hyperfluorescence \\
        OCTA &  hyperfluorescent neovascular network\\ 
    \end{tabular}
\end{table}

\begin{figure}
    \centering
    \includegraphics[width=0.155\textwidth]{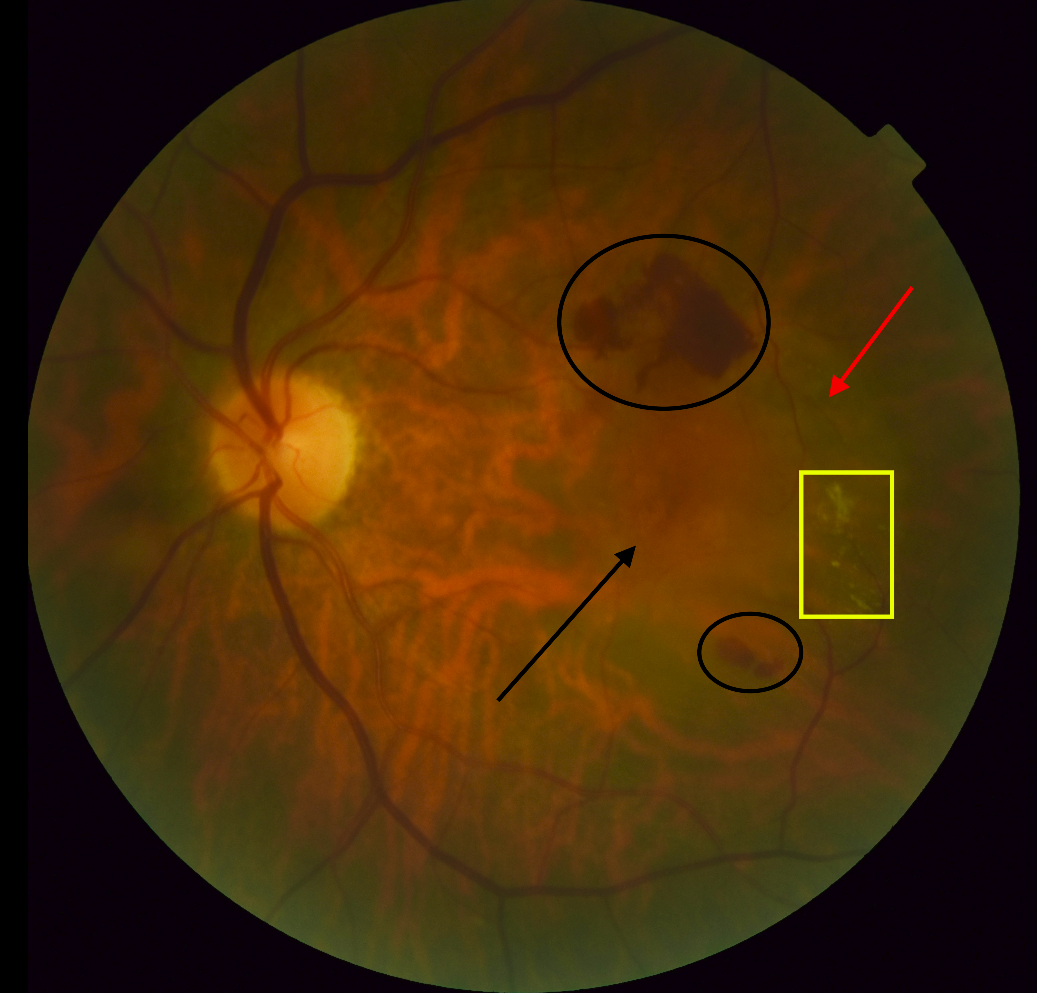}  \hfill
    \includegraphics[width=0.15\textwidth]{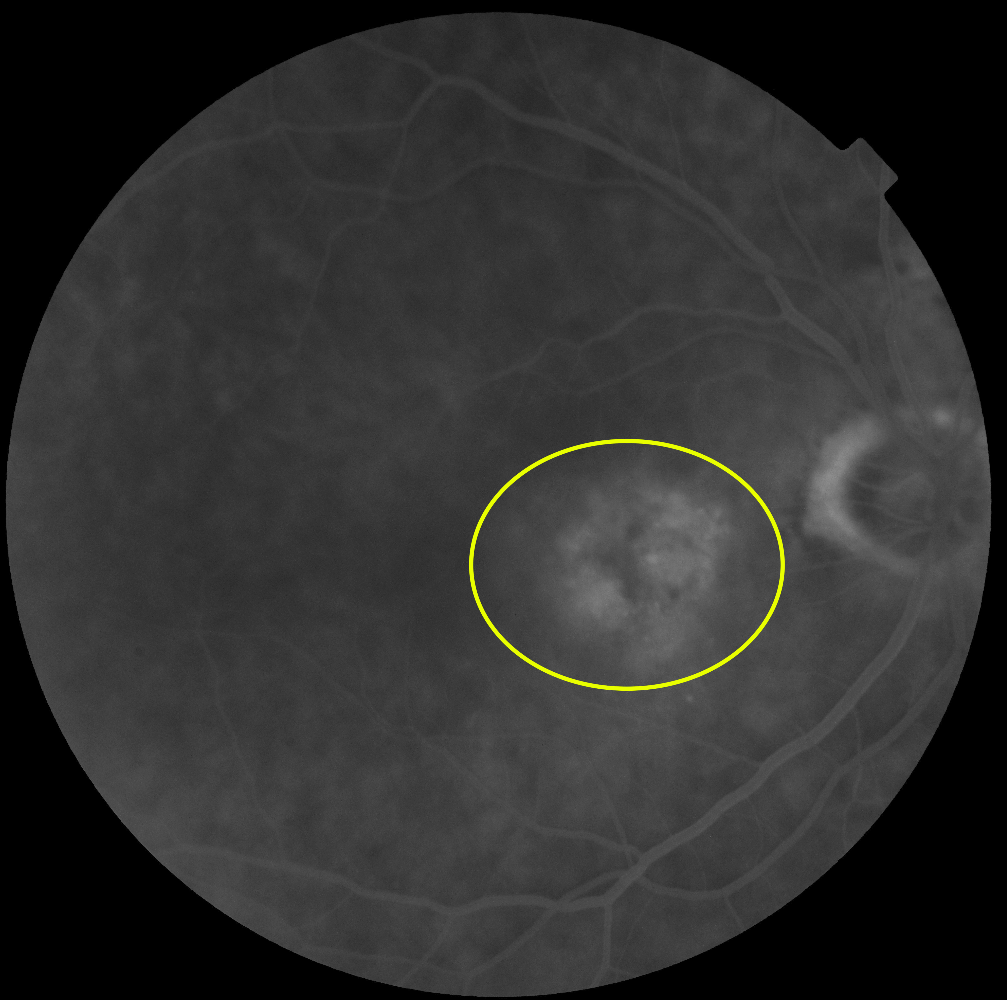}  \hfill
    \includegraphics[width=0.15\textwidth]{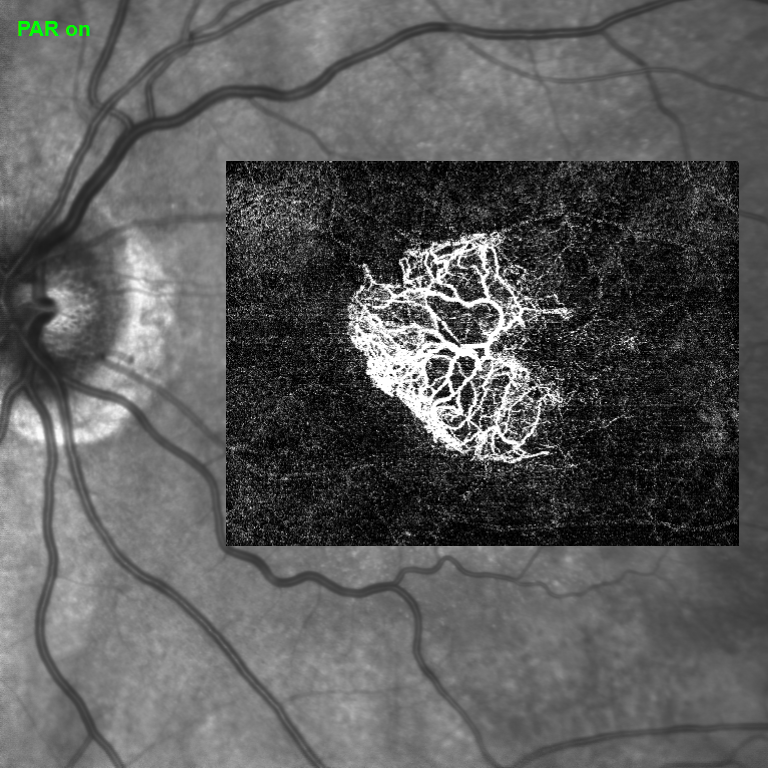} \hfill
    \caption{Color fundus photography (left) reveals hemorrhages (black circles), hard exudates (yellow rectangle), subretinal fluid (red arrow) and RPE detachment (black arrow). FA (center) reveals in the selected frame the speckled hyperfluorescence (yellow circle). 
    OCTA (right) reveals the neovascular membrane as a hyperfluorescent neovascular network.}
    \label{fig:hemo}
\end{figure}

\begin{figure}
    \centering
    \includegraphics[width=0.38\textwidth]{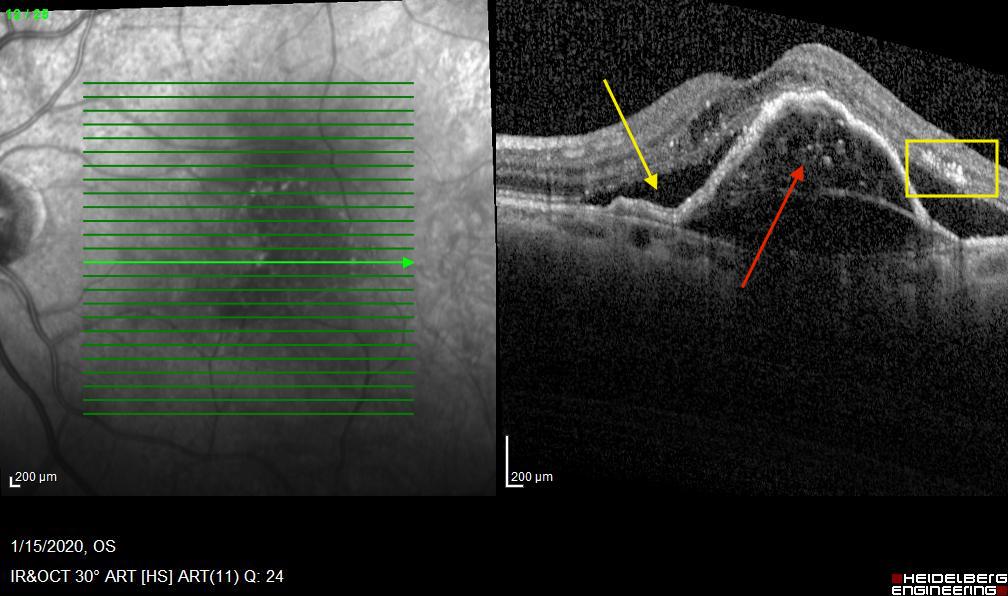}
    \caption{OCT reveals elevation of the RPE (red arrow), subretinal fluid (yellow arrow) and hard exudates (yellow rectangle)}
    \label{fig:oct}
\end{figure}

The ensemble of image techniques cover different perspectives and retina biomarkers (Table~\ref{tab:img}).
Color fundus photography reveals hemorrhages (Figure~\ref{fig:hemo} left). 
Fluorescein angiography (FA) reveals speckled hyperfluorescence (Figure~\ref{fig:hemo} center), while 
 OCTA sees the neovascular membrane as a hyperfluorescent neovascular network (Figure~\ref{fig:hemo} right).
The corresponding OCT scan for Figure~\ref{fig:hemo}) better shows biomarkers like elevation of the retinal pigment epithelium, subretinal fluid or hard exudates (Figure~\ref{fig:oct}). 

Five OCTA-criteria could be used to describe a neovascular lesion in AMD~\cite{wang2021optical}:
(1) tiny branching vessels (thin, tangled capillaries) vs. mature vessels (linear filamentous capillaries); 
(2) anastomotic arcade (peripheral connection, lacy wheel appearance); 
(3) inner loops (inner anastomoses); 
(4) perilesional hypointense halo (hypointense area around NV, local choriocapillaris alteration);
(5) feeder vessel.
For the quantitative assessment, parameters such as mCNV area ($mm^2$), vessel area ($mm^2$), vessel density, vessel length (mm), vessel diameter ($\mu$m), fractal dimension, vessel junctions, junction density (n/mm), and vessel tortuosity can be calculated.

First, we focus on identifying these biomarkers for nAMD in OCTA images. Second, we develop a 3D visualization of the neovascularization. Third, we experiment with various white-box learning algorithms to enhance explainability, and we build an ensemble of classifiers to improve robustness.

\section{Creating the OCTA dataset}

\begin{table*}
    \centering  
    \caption{Available OCTA datasets}
    \label{tab:dataset_info}
    \begin{tabular}{llllll}
        \textbf{Dataset} & \textbf{Dataset Size} & \textbf{Acquisition device} & \textbf{Resolution} & \textbf{Access} & \textbf{No.} \\
        \hline
        ROSE-1 & 117 B-scan & Optovue & 304$\times$304 & open & 2 \\
        ROSE-2 & 112 B-scan & Heidelberg OCT2 & 512$\times$512 & open & 1 \\
        \href{https://ieee-dataport.org/documents/ss-oct-sa}{SS-OCT SA} & 73 volumes & Artificially synthesized & - & open & - \\
        \href{https://ieee-dataport.org/documents/octdl-optical-coherence-tomography-dataset-image-based-deep-learning-methods}{OCTDL} & 2064 B-scan & Optovue Avanti & - & open & 6 \\
        \href{https://people.duke.edu/~sf59/RPEDC_Ophth_2013_dataset.htm}{SDOCT} & 35400 B-scan & Bioptigen & - & open & 1 \\
        \href{https://ieee-dataport.org/open-access/octa-500}{OCTA-500} & 361600 B-scan & RTVue-XR, Oprovue & 304$\times$304 & limited & $>$12 \\
        RETOUCH & 70 volumes & Citrrus, Spectralis, Topcon & 512$\times$1024 & limited & 2 \\
        FAZID & 304 B-scan & not mentioned & 420$\times$420 & limited & 3 \\
        OCTAGON & 108 B-scan & not mentioned & 320$\times$320 & limited & 4 \\
    \end{tabular}
\end{table*}

\begin{table}
    \centering
    \caption{Batches of the collected dataset}
    \label{tab:batch_info}
    \begin{tabular}{crl}
        \textbf{Batch}  & \textbf{Img.}  & \textbf{Description} \\ \hline
        1  &  124 & Clear view of the blood vessel \\
        2  &  98  & Sections of the same patient \\
        3  &  124 & Annotated dataset for testing \\
        4  &   65 & Healthy retina examples \\
    \end{tabular}
\end{table}

Since OCTA is relatively new, the available technical instrumentation (e.g., datasets, code) is concentrated in specific, narrow types of image analysis. Kylyabin et al. provide a detailed analysis of ocular disease datasets~\cite{kulyabin2024octdl}, and a collection of open tools for retinal conditions is available at \href{https://people.duke.edu/~sf59/software.html}{Duke University}. The existing datasets are presented in Table~\ref{tab:dataset_info}. 
Datasets such as SDOCT~\cite{farsiu2014quantitative}, ROSE-1, and ROSE-2~\cite{ma2020rose} focus on at most two conditions, none of which are related to nAMD. The SS-OCT SA dataset is artificially generated. RETOUCH~\cite{8653407} and OCTA-500~\cite{li2024octa} use different types of equipment for image acquisition, and the format and color spectrum do not align with our preprocessing steps, which are adjusted for the Heidelberg OCTA format. The FAZID~\cite{agarwal2020foveal} and OCTAGON datasets present OCT scans.

\begin{figure}
    \centering
    \includegraphics[width=0.15\textwidth]{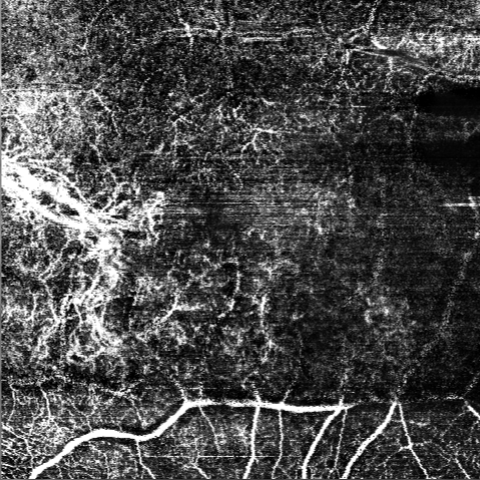}
    \includegraphics[width=0.15\textwidth]{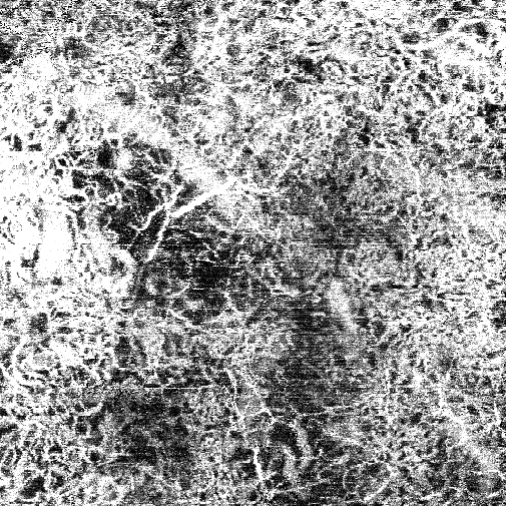}
    \includegraphics[width=0.16\textwidth]{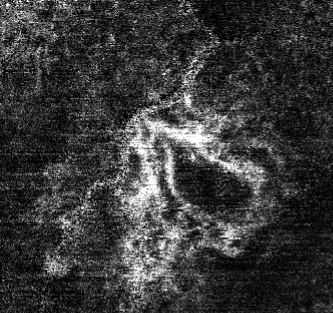}
    \caption{Example of images removed from dataset because of \\ (left) artifacts (middle), high exposure (right), sand appearance}
    \label{fig:notgood_batch1}
\end{figure}
We collected 304 anonymized grayscale images from the Heidelberg OCTA device~\cite{soomro2018use} used at the Clinic of Ophthalmology, Emergency County Hospital, Cluj-Napoca. The device can capture around 512 images within the retina during a single scan. The dataset is divided in four batches (right part in Table~\ref{tab:dataset_info}).
\textit{The first batch} initially contained 133 OCTA images. 
Artifacts and abnormal blood vessels (Fig.~\ref{fig:notgood_batch1}) led to 8 images being eliminated 
\textit{The second batch} consists in 98 OCTA images from the same patient. 
Among them, 46 images show multiple layers in the left eye, and 50 images show the right eye.
\textit{The third batch} includes only images used for measuring the performance of our solution. 
\textit{The fourth batch} consists in 73 images of healthy retina from different patients.

\section{Extracting Biomarker from OCTA} 

\subsection{Processing workflow}
\begin{figure*}
    \centering
    \includegraphics[width=0.9\textwidth]{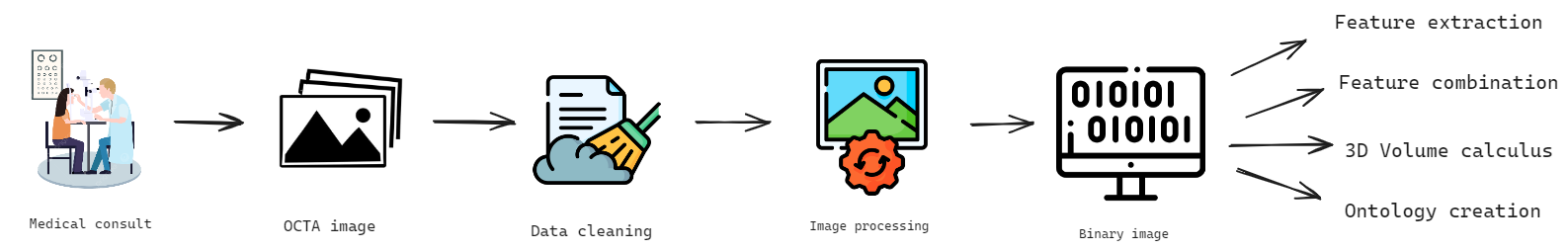}
    \caption{System architecture}
    \label{fig:sysarch}
\end{figure*}

The workflow (Fig.~\ref{fig:sysarch}) consists in a fixed number of 3 phases, that provide the metrics needed for further calculations or development. The metrics are extracted of the images based on the calculation of the number of object pixels. Among the metrics we mention \textit{mCNV Area, Vessel Area, Vessel Density}.

After the medical consultation, the grayscale image is saved and then goes through a \textit{Data Cleaning} phase, consisting of cropping out unnecessary parts of the photo and applying a Gaussian filter.

In the \textit{Image Processing} phase, different kernels and operations are applied to the photo, including Otsu Thresholding, Multiple and Binary Thresholding, Salt and Pepper filtering, and selecting the main component. The output is a binary image that enhances the blood vessels present at the retina level. From this point, multiple operations can be performed on these images. Among them are Feature Extraction, Feature Combination, 3D Volume Calculations, and Ontology creation. 
The code is available at \href{https://github.com/BBsimonaBB/Measuring-nAMD-biomarkers-from-OCTA-images}{repository}, also as Docker images.

\subsection{Data Cleaning}

\begin{figure}
    \centering
    \includegraphics[width=0.45\textwidth]{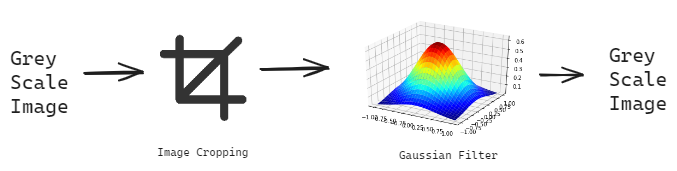}
    \caption{Data Cleaning}
    \label{fig:dc}
\end{figure}

The \textit{Data Cleaning} phase (Fig.~\ref{fig:dc}) consists of two steps: image cropping and Gaussian filtering for smoothing. In reality, the blood vessel is typically represented as a single piece. However, the process of capturing the OCTA image tends to distort some images, making a vessel appear fragmented, even though it is actually a single continuous thread. Without this phase, relevant information could be lost, or simple artifacts might be misinterpreted as blood vessels.

\begin{figure}
    \centering
    \includegraphics[width=0.18\textwidth]{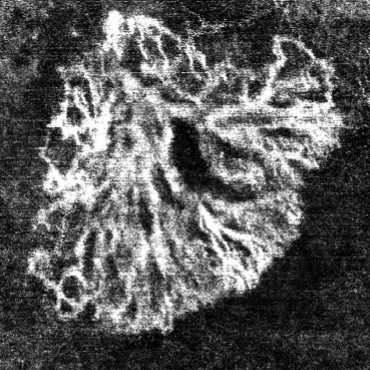}
    \includegraphics[width=0.18\textwidth]{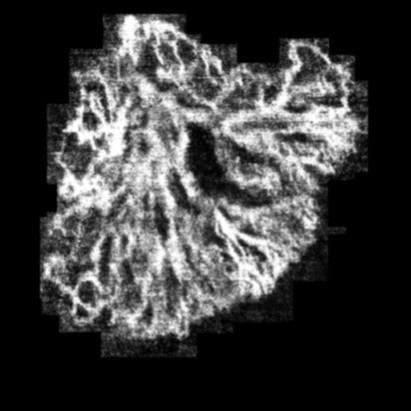}
    \caption{Data cleaning (white is the spreaded blood vessel causing nAMD): (left) before; (right) after data cleaning}
    \label{fig:example_clean_1}
\end{figure}

In Fig.~\ref{fig:example_clean_1}, one can observe that the input image has a granular structure, causing the object pixels to appear disconnected. A significant number of background pixels interfere with the visualization of the blood vessels.

\subsection{Image Processing}
The image entering this phase is a grey scale one. On the resulting image, a binary one, classic image processing techniques applied.

\begin{figure}
    \centering
    \includegraphics[width=0.49\textwidth]{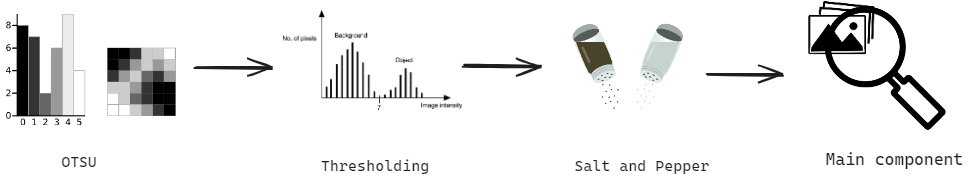}
    \caption{Image Processing}
    \label{fig:pro}
\end{figure}

The \textit{Image Processing} phase (Fig.~\ref{fig:pro}) includes 4 steps: Otsu Thresholding, Multiple and Binary Thresholding, as the main thresholding techniques, followed by Salt and Pepper Filtering and Selecting the Main Component. 
All these steps use a kernel of fixed size and iterate over all the pixels.

First, Otsu's thresholding~\cite{khairnar2021effect}, an image segmentation technique used to automatically determine an optimal threshold value for converting a grayscale image into a binary image is applied. Normally, Otsu's method is used for generating a binary image. 
However, the output of the current solution is still a grayscale image. The upper thresholding limit is set to 170 instead of 255.
This adjustment is useful in further processing to distinguish among different types of highlights.

Second, Multiple Thresholding~\cite{mittal2018optimum} and Binary Thresholding are applied. 
Multiple Thresholding applies several threshold values to segment an image into multiple levels or bands of grayscale. 
Binary thresholding is applied afterward to simplify each of these bands into a binary format.

Third, a Salt and Pepper filter~\cite{thanh2020two} aimed to eliminate fine noise that resembles grains of salt or pepper.

Fourth, largest component in the image is selected, specifically \textit{Component Labeling}~\cite{safaei2022automatic}. After labeling, we keep only the largest component, which contains the most object pixels.

\subsection{Feature Extraction}
Blood vessels in the back of the eye are so small they often grow or shrink with just a few microns in time. Specialists have defined 9 biomarkers to measure and quantify the quantity of blood present at that level in the retina. 
Biomarkers stand for means of evaluating the progression and severity of nAMD. The ones like mCNV Area and Vessel Area can be directly extracted from images, while others, such as Vessel Density, are calculated based on the combination of others. 

We are interested here in two biomarkers: the \textit{mCNV Area} and the \textit{Total Area}. The \textit{mCNV Area} is the area of all the object pixels (i.e., white ones). This is computed by calculating the area of a single pixel and multiplying it by the number of object pixels in that image. 

The \textit{Total Area} feature is calculated similarly, by counting the number of object pixels. However, the image being analyzed is first processed through a Region Fill operation, which aims to fill in the black holes within the initial blood vessels.
The area of a pixel is calculated with~\cite{dougherty2020digital}: 
\begin{equation}
\mathcal{A}_{pixel}=\left( \frac{\text{Size of the image} (\mu m)}{\text{Number of pixels}} \times \frac{1 \text{ mm}}{1000 \mu m} \right)^2
\end{equation}

\begin{figure}
    \centering
    \includegraphics[width=0.2\textwidth]{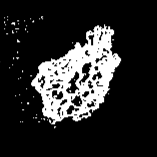}
    \includegraphics[width=0.2\textwidth]{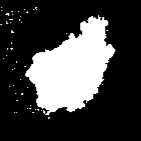}
    \caption{OCTA image inside a patient's retina, after image processing, meant to highlight the blood vessels (white pixels): (left) Binary Image (right) Binary Image after region filling}
    \label{fig:features}
\end{figure}

Knowing that an OCTA image has a scale of $200\mu m \times 200\mu m$ and that each OCTA image contains $510\times510$ pixels, we find the size of one pixel to be 0.154 $\mu m^2$.
The original image after the "Image Processing" phase focuses on the blood vessel area, represented by the white pixels. It indicates the size of the active mCNV lesion. The internal black areas show parts of the object where the blood vessel is missing. Figure~\ref{fig:features} illustrates a structure similar to a sponge.

To calculate the total area, we need to pass the output image from the "Image Processing" phase through another operation, \textit{Region Fill}, to fill in the black missing sections from the initial binary image. After applying this filter, the total vessel area will be larger than the mCNV area:
$
\mathcal{A}_{Total} \geq \mathcal{A}_{mCNV}$.

Feature combination is then applied to compute the \textit{Vessel Density}, which is the ratio of the \textit{mCNV Area} to the \textit{Total Area}:
$VesselDensity= \frac{\mathcal{A}_{mCNV} }{\mathcal{A}_{Total}}$.

\subsection{Running experiments}
\begin{figure*}
    \centering
    \includegraphics[width=0.17\textwidth]{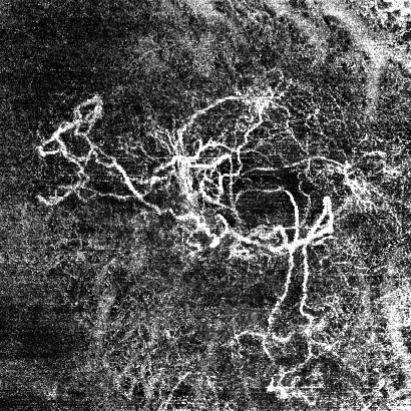} \hfill
    \includegraphics[width=0.17\textwidth]{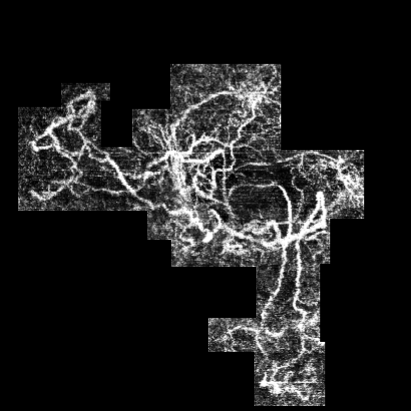}\hfill
    \includegraphics[width=0.17\textwidth]{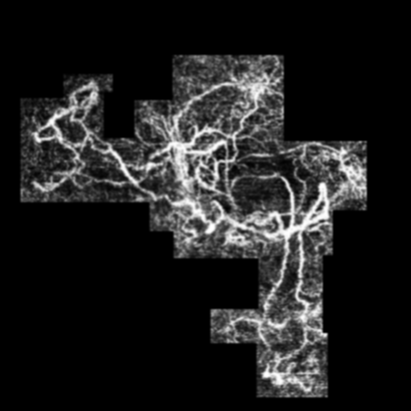} \hfill
    \includegraphics[width=0.17\textwidth]{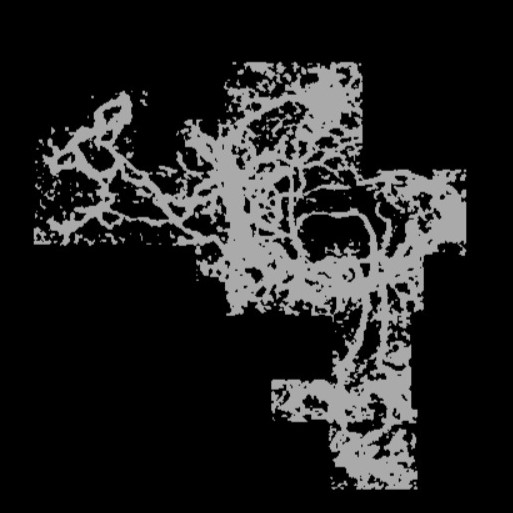}
    \caption{(a) Original image; (b) After black box cropping; (c) Applying Gaussian filter; (d) Applying OTSU filter}
    \label{fig:processing}
\end{figure*}

We illustrate here how the processing steps lead to the binary image, which contains essential information about the blood vessels.
Fig.~\ref{fig:processing} shows the steps involved in the Data Cleaning phase and Otsu Thresholding. 

\begin{figure*}
    \centering
    \includegraphics[width=0.17\textwidth]{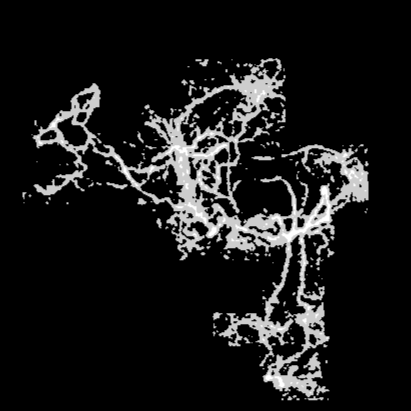}\hfill
    \includegraphics[width=0.17\textwidth]{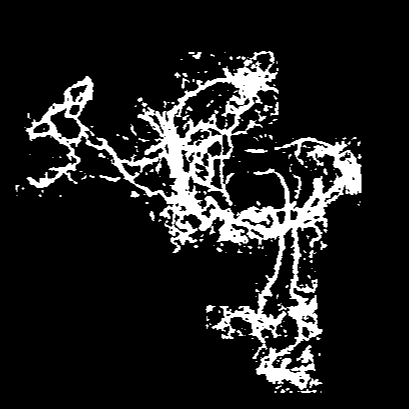}\hfill
    \includegraphics[width=0.17\textwidth]{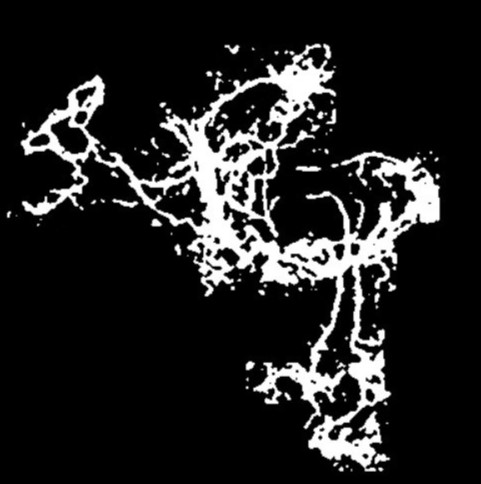}\hfill
    \includegraphics[width=0.17\textwidth]{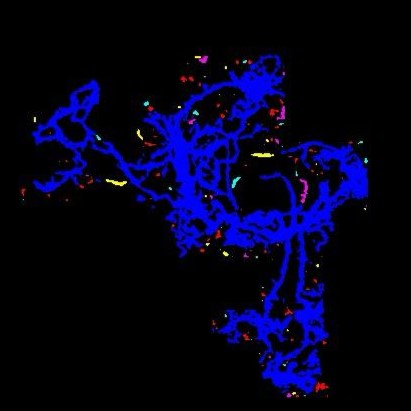}
    \caption{(a) Multiple binarization; (b) Automated binarization; (c) 
    Applying salt and pepper filter; (d) All resulting components}
    \label{fig:processing2}
\end{figure*}

\begin{figure}
    \centering
    \includegraphics[width=0.17\textwidth]{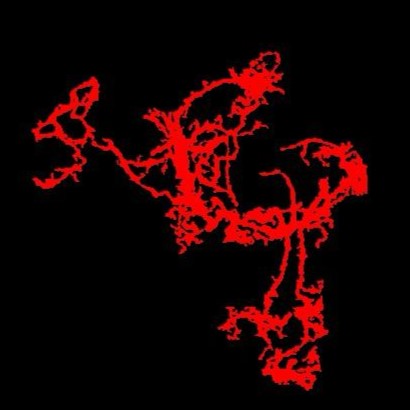}\hfill
    \includegraphics[width=0.17\textwidth]{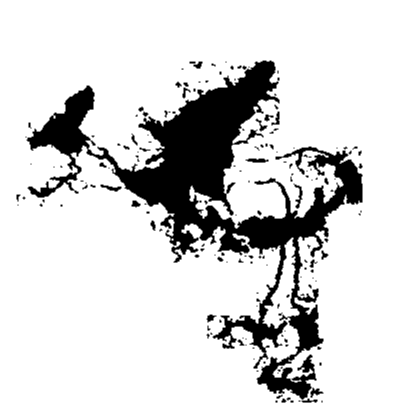}
    \caption{Main component and region fill}
    \label{fig:result}
\end{figure}

Fig.~\ref{fig:processing2} presents the image processing techniques and their outcomes at each iteration. 
The grayscale images progressively approach a binary representation, as seen after the application of multiple binarization techniques. 
The salt and pepper filter removes part of the noise. 
In the final image, all components are colored in a way that highlights the principal component. 
Any remaining noise is removed in Fig.~\ref{fig:result}, which also shows the result after applying region filling techniques.

To assess the performance, we used the Jaccard Distance = $1- \frac{|A \cap B|}{|A \cup B|}$ and the Dice coefficient  $\frac{2 |A \cap B|}{|A| + |B|}$ to compare the two sets of data in our scenario: the 1st batch and the 3rd one. 
A Jaccard Index is 0.8912, suggesting that 89.12\% of the pixels identified during the Data Cleaning and Image processing module were also identified by the ophthalmologist. 
The Dice Coefficient ranges from 0 to 1, where 0 indicates no overlap, and 1 indicates complete overlap (identical sets). The value of the Dice coef = 0.92 is very high, which suggests that the two sets A and B have a large overlap.
\begin{figure}
    \centering
    \includegraphics[width=0.2\textwidth]{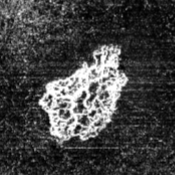}
    \includegraphics[width=0.15\textwidth]{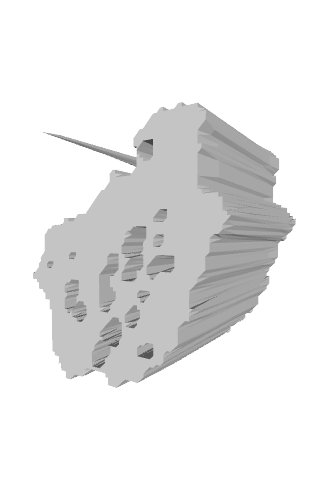}
    \includegraphics[width=0.2\textwidth]{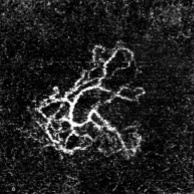}
    \includegraphics[width=0.15\textwidth]{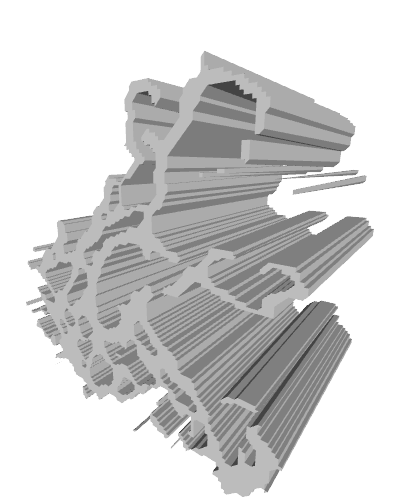}
    \caption{(a) First visit in OCTA format (b) First visit in STL format (c) Second visit, after treatment in OCTA format (d) Second visit, after treatment in STL format}
    \label{fig:visits}
\end{figure}

The pictures represent the right eye of a patent suffering from nAMD. The first image - (a) from Fig. ~\ref{fig:visits}  - was taken on June 27th 2023 and the second one - (c) from Fig. ~\ref{fig:visits}  - was taken almost a year later, on June 11th 2024.

\begin{table}
    \centering
    \caption{Measurements}
    \label{tab:volume}
    \begin{tabular}{c|c|c|r|c}
        \textbf{Visit} & \textbf{Sections} & \textbf{Distance (\(\mu\text{m}\))} & \textbf{Pixels} & \textbf{Volume (\(\mu\text{m}^3\))} \\ \hline
        27.07.23 & 50 & 25  & 112.037 & 469842.5 \\
        11.06.24 & 52 & 25  & 93.112  & 340459.4 \\ \hline
    \end{tabular}
\end{table}

\section{White Box Machine Learning}

From a clinical perspective, white box machine learning models provide an interpretable framework that enhances trust and acceptance. From a technical perspective, the datasets available in clinical practice are often not large enough to support deep learning algorithms. Moreover, we have already extracted the features of interest as numerical values. Thus, machine learning (ML) algorithms like decision trees or support vector machines (SVM) are more appropriate.

To detect the nAMD condition based on the extracted biomarkers, we applied three white box learning algorithms. First, we extracted rules from a decision tree. Second, we extracted feature vectors learned using SVM. 
Third, we employed an inductive logic programming method, namely DL-Learner, to learn axioms in Description Logic related to the nAMD condition. Finally, the rules extracted from these three methods were combined in an ensemble of classifiers.

\begin{table}
    \centering
    \caption{Decision Tree Rules for mCNV and Vessel Area}
    \label{tab:decision_tree_rules}
    \begin{tabular}{lrl}
        \textbf{Rule} & \textbf{Label} & \textbf{Rule} \\ \hline
        $R^{DT}_1$ & $\neg$Sick & mCNV $\leq$ 0.01 \\
        $R^{DT}_2$ & Sick & mCNV $>$ 0.01 $\land$ vessel $\leq$ 0.02 \\
        $R^{DT}_3$ & $\neg$Sick & mCNV $>$ 0.01 $\land$ vessel $>$ 0.02 $\land$ mCNV $\leq$ 0.03 \\
        $R^{DT}_4$ & Sick & mCNV $>$ 0.01 $\land$ vessel $>$ 0.02 $\land$ mCNV $>$ 0.03 \\ \hline
    \end{tabular}
\end{table}
\begin{table}
    \centering
    \caption{Rules generated by the SVM}
    \begin{tabular}{lrl}
        \textbf{Rule} & \textbf{Label} & \textbf{Rule} \\
        \hline
        $R^{SVM}_1$ & Sick & $M_{sm} \leq .5 \wedge M_{med} \leq .5$ \\
        $R^{SVM}_2$ & $\neg$ Sick & $M_{sm} \leq .5 \wedge M_{med} > .5 \wedge V_{med} < .5 \wedge V_{sm} < .5$ \\
        $R^{SVM}_3$ & $\neg$ Sick & $M_{sm} \leq .5 \wedge M_{med} > .5 \wedge V_{med} < .5 \wedge V_{sm} < .5$ \\
    \end{tabular}
    \label{tab:svm_rules}
\end{table}

First, the \textit{learning rules from the decision tree} (Table~\ref{tab:decision_tree_rules}) achieved a testing accuracy of 0.68. 
Second, the \textit{learning rules from SVM} using the RBF kernel generated 3 rules, with an accuracy of 0.60, (Table~\ref{tab:svm_rules}). 
Third, \textit{learning axioms in description logic} is based on an ontology formalizing the nAMD condition. The small nAMD ontology that we have engineered is a formal representation of knowledge related to nAMD, structured around three elements: 
(i) classes or concepts, (ii) properties and relationships between these concepts, and (iii) individuals or instances of these concepts. 
For engineering the small nAMD ontology, we used the Protege ontology editor, while the learning process was performed using the DL-Learner tool. 
JSON rules were employed to link the extracted features with the triple-based format required by the DL-Learner algorithms.

At the top level, there are two disjoint classes: $Sick \sqsubseteq \neg NotSick$. 
There are also two main object properties: the mCNV area and the vessel area, both of which are categorized into three labels: \(small, medium, big\). 
The individuals are represented by images, with each image name used as its identifier. Each image has a $totalArea$ and an $mCNVArea$ as data properties. 
There are 65 images representing healthy individuals and 120 images from individuals with nAMD.

Based on the learned axioms, the Hermit reasoning engine was used to classify new images~\cite{abicht2023owl}. 
We generated equivalence axioms and subclass axioms for the two classes: \(Sick, NotSick\). 
All expressions achieved 100 percent accuracy, suggesting that each class expression perfectly fits the data used for learning. 
This could indicate either a well-defined dataset with clear boundaries based on the provided conditions, or a potential case of overfitting, where the model fits exactly to the training data but might not generalize (Table~\ref{tab:dl_learner_rules_sick}).

\begin{table}
    \centering
    \caption{DL-Learner rules for the Sick Class}
    \begin{tabular}{ll}
        \textbf{Generated Rules} & \textbf{Redefined Rules} \\
        \hline
        (Sick and mCNVArea) or (v\_big and v\_medium) & $V_{big} \wedge V_{med}$ \\
        (Sick and VesselArea) or (m\_big and v\_medium) & $M_{big} \wedge V_{med}$ \\
        (Sick and VesselArea) or (m\_small and v\_medium) & $M_{small} \wedge V_{med}$ \\
        (Sick and mCNVArea) or (v\_medium and v\_small) & $V_{med} \wedge V_{small}$ \\
        (Sick and mCNVArea) or (m\_small and v\_medium) & $M_{small} \wedge V_{med}$ \\
        Image and (Sick or (m\_small and v\_medium)) & $M_{small} \wedge V_{med}$ \\
    \end{tabular}
    \label{tab:dl_learner_rules_sick}
\end{table}

The learned models are small and transparent. Based on DT we obtained 4 rules that provide 0.65\% accuracy.
Based on SVM we extracted 3 rules (0.65\% accuracy), while with DL-learner, 10 axioms in description logics were generated.


To improving performance with ensemble learning, the data was discretized using a supervised method. More specifically, the numerical information was split in 3 categories, \(small, medium, big\). The \(m small\) stands for mCNV and \(v small\) for vessel area. The supervised discretization was done having as target the \(Sick, NotSick\) column.
The images were split in train and test categories with the ratio of 80/20. 

\section{Discussion and Related Work}
OCTA images have been used to extract
 nine biomarkers  of Myopic Choroidal neovascularization (mCNV)~\cite{deshpande2023imagej}:
(1) mCNV area
(2) vessel area 
(3) vessel density 
(4) vessel length 
(5) vessel diameter 
(6) vessel junctions 
(7) junction density 
 (8) fractal dimension 
(9) vessel tortuosity. 
The limitations of this implementation refer to relying solely on images obtained from Heidelberg OCTA device. Because of differences such as in image format, contrast, overall structure of the visualizations of the OCTA images, the current implementation is difficult to be used on datasets obtained from other OCTA devices.


Related projects, have shown high overlap metrics—such as a Dice coefficient of 0.86 and Jaccard index of 0.82~\cite{compare1}, and another achieving a Dice coefficient of 0.89 with a Jaccard index of 0.79~\cite{compare2}. 
We obtained a Jaccard index of 0.8912,  and a Dice coefficient of 0.92, reflecting a high level of overlap between the automated results and expert assessment. 

\textit{Human in the loop} occurs when cropping the images during Data Cleaning. The ophthalmologist distinguishes between artifacts and blood vessels and can manually analyze the rules generated by the DT, SVM, or DL-Learner. 
For instance, the original rule: $(Sick \land mCNVArea) \lor (V_{med} \land V_{small}) \rightarrow nAMD$ could be simplified to: $V_{med} \land V_{small} \rightarrow nAMD$. In the same line with~\cite{groza2021agents}, the human expert has the possibility to augment the learned models with own expert knowledge. 

The mainstream of AI in medical diagnosis typically involves feeding images into deep learning systems. In contrast, our approach is: (1) to identify biomarkers as they appear in medical protocols or clinical practice, and 
(2) to apply white box machine learning on the extracted numerical features. 
These white-box models offer several advantages in ophthalmology:
(i) clinicians can understand the decision-making process, which enhances trust and acceptance of these tools in clinical practice; 
(ii) regulatory bodies are more likely to approve such white-box models;
(iii) ophthalmologists can identify new biomarkers or patterns in ophthalmic data; 
(iv) residents can use white-box algorithms as educational tools to improve their skills.

\section{Conclusion}
The handle various aspects of the nAMD condition, there is a current need for transparent support tools for ophthalmologist.  
To support decisions of physicians, we developed here three instruments: (i) extracting biomarkers specific to nAMD form OCTA images, (ii) providing a 3D visualisation of the neovascularisation; (iii) applying a white box machine learning ensemble to classify images and to show the physicians the rules used in that classification.      

\bibliographystyle{IEEEtran}
\bibliography{bib}

@article{deshpande2023imagej,
  title={An ImageJ macro tool for OCTA-based quantitative analysis of Myopic Choroidal neovascularization},
  author={Deshpande, Aadit and Raman, Sundaresan and Dubey, Amber and Susvar, Pradeep and Raman, Rajiv},
  journal={Plos one},
  volume={18},
  number={4},
  pages={e0283929},
  year={2023},
  publisher={Public Library of Science San Francisco, CA USA}
}

@article{li2024octa,
  title={OCTA-500: a retinal dataset for optical coherence tomography angiography study},
  author={Li, Mingchao and Huang, Kun and Xu, Qiuzhuo and Yang, Jiadong and Zhang, Yuhan and Ji, Zexuan and Xie, Keren and Yuan, Songtao and Liu, Qinghuai and Chen, Qiang},
  journal={Medical Image Analysis},
  volume={93},
  pages={103092},
  year={2024},
  publisher={Elsevier}
}

@Article{diag2023,
AUTHOR = {Muntean, G.A. and Marginean, A. and Groza, A. and Damian, I. and Roman, S.A. and Hapca, M.C. and Muntean, M.V. and Nicoară, S.D},
TITLE = {The Predictive Capabilities of Artificial Intelligence-Based OCT Analysis for Age-Related Macular Degeneration Progression—A Systematic Review},
JOURNAL = {Diagnostics},
VOLUME = {13},
YEAR = {2023},
NUMBER = {14},
ARTICLE-NUMBER = {2464},
URL = {https://www.mdpi.com/2075-4418/13/14/2464},
ISSN = {},
DOI = {10.3390/diagnostics13142464},
type ={journal},
infosite = {https://www.mdpi.com/2075-4418/13/14/2464},
durl = {https://www.mdpi.com/2075-4418/13/14/2464/pdf?version=1690265761}
}

@article{wang2021optical,
  title={Optical coherence tomography angiography-based quantitative assessment of morphologic changes in active myopic choroidal neovascularization during anti-vascular endothelial growth factor therapy},
  author={Wang, Yao and Hu, Zhongli and Zhu, Tiepei and Su, Zhitao and Fang, Xiaoyun and Lin, Jijian and Chen, Zhiqing and Su, Zhaoan and Ye, Panpan and Ma, Jian and others},
  journal={Frontiers in Medicine},
  volume={8},
  pages={657772},
  year={2021},
  publisher={Frontiers Media SA}
}

@article{liu2015automated,
  title={Automated choroidal neovascularization detection algorithm for optical coherence tomography angiography},
  author={Liu, Li and Gao, Simon S and Bailey, Steven T and Huang, David and Li, Dengwang and Jia, Yali},
  journal={Biomedical optics express},
  volume={6},
  number={9},
  pages={3564--3576},
  year={2015},
  publisher={Optica Publishing Group}
}

@article{de2015review,
  title={A review of optical coherence tomography angiography (OCTA)},
  author={De Carlo, Talisa E and Romano, Andre and Waheed, Nadia K and Duker, Jay S},
  journal={International journal of retina and vitreous},
  volume={1},
  pages={1--15},
  year={2015},
  publisher={Springer}
}

@article{soomro2018use,
  title={The use of optical coherence tomography angiography for detecting choroidal neovascularization, compared to standard multimodal imaging},
  author={Soomro, Taha and Talks, James},
  journal={Eye},
  volume={32},
  number={4},
  pages={661--672},
  year={2018},
  publisher={Nature Publishing Group}
}

@book{dougherty2020digital,
  title={Digital image processing methods},
  author={Dougherty, Edward R},
  year={2020},
  publisher={CRC Press}
}

@article{abicht2023owl,
  title={OWL Reasoners still useable in 2023},
  author={Abicht, Konrad},
  journal={arXiv preprint arXiv:2309.06888},
  year={2023}
}

@article{khairnar2021effect,
  title={Effect of image binarization thresholds on breast cancer identification in mammography images using OTSU, Niblack, Burnsen, Thepade's SBTC},
  author={Khairnar, Smita and Thepade, Sudeep D and Gite, Shilpa},
  journal={Intelligent Systems with Applications},
  volume={10},
  pages={200046},
  year={2021},
  publisher={Elsevier}
}

@article{mittal2018optimum,
  title={An optimum multi-level image thresholding segmentation using non-local means 2D histogram and exponential Kbest gravitational search algorithm},
  author={Mittal, Himanshu and Saraswat, Mukesh},
  journal={Engineering Applications of Artificial Intelligence},
  volume={71},
  pages={226--235},
  year={2018},
  publisher={Elsevier}
}

@article{thanh2020two,
  title={A two-stage filter for high density salt and pepper denoising},
  author={Thanh, Dang NH and Hai, Nguyen Hoang and Prasath, Surya and Hieu, Le Minh and Tavares, Jo{\~a}o Manuel RS},
  journal={Multimedia tools and applications},
  volume={79},
  number={29},
  pages={21013--21035},
  year={2020},
  publisher={Springer}
}

@article{safaei2022automatic,
  title={An automatic image processing algorithm based on crack pixel density for pavement crack detection and classification},
  author={Safaei, Nima and Smadi, Omar and Masoud, Arezoo and Safaei, Babak},
  journal={International Journal of Pavement Research and Technology},
  volume={15},
  number={1},
  pages={159--172},
  year={2022},
  publisher={Springer}
}

@article{kulyabin2024octdl,
  title={OCTDL: Optical Coherence Tomography Dataset for Image-Based Deep Learning Methods},
  author={Kulyabin, Mikhail and Zhdanov, Aleksei and Nikiforova, Anastasia and Stepichev, Andrey 
          and Kuznetsova, Anna and et al.},
  journal={Scientific Data},
  volume={11},
  number={1},
  pages={365},
  year={2024},
  publisher={Nature Publishing Group UK London},
  doi={https://doi.org/10.1038/s41597-024-03182-7}
}

@article{farsiu2014quantitative,
  title={Quantitative classification of eyes with and without intermediate age-related macular degeneration using optical coherence tomography},
  author={Farsiu, Sina and Chiu, Stephanie J and O'Connell, Rachelle V and Folgar, Francisco A and Yuan, Eric and Izatt, Joseph A and Toth, Cynthia A and Age-Related Eye Disease Study 2 Ancillary Spectral Domain Optical Coherence Tomography Study Group and others},
  journal={Ophthalmology},
  volume={121},
  number={1},
  pages={162--172},
  year={2014},
  publisher={Elsevier}
}

@article{groza2021agents,
title={Agents that argue and explain classifications of retinal conditions},
author={Groza, Adrian and Toderean, Liana and Muntean, George Adrian and Nicoara, Simona Delia},
journal={Journal of Medical and Biological Engineering},
volume={41},
number={5},
pages={730--741},
year={2021},
publisher={Springer},
doi ={10.1007/s40846-021-00647-7},
type = {journal},
durl = {https://www.researchgate.net/publication/352268728_Agents_that_argue_and_explain_classifications_of_retinal_conditions}
}

@article{ma2020rose,
  title={ROSE: a retinal OCT-angiography vessel segmentation dataset and new model},
  author={Ma, Yuhui and Hao, Huaying and Xie, Jianyang and Fu, Huazhu and Zhang, Jiong and Yang, Jianlong and Wang, Zhen and Liu, Jiang and Zheng, Yalin and Zhao, Yitian},
  journal={IEEE transactions on medical imaging},
  volume={40},
  number={3},
  pages={928--939},
  year={2020},
  publisher={IEEE}
}

@inproceedings{agarwal2020foveal,
  title={The foveal avascular zone image database (fazid)},
  author={Agarwal, Arpit and Raman, Rajiv and Lakshminarayanan, Vasudevan and others},
  booktitle={Applications of Digital Image Processing XLIII},
  volume={11510},
  pages={507--512},
  year={2020},
  organization={SPIE}
}

@ARTICLE{8653407,
  author={Bogunović, Hrvoje and Venhuizen, Freerk and Klimscha, Sophie and Apostolopoulos, Stefanos and et al.},
  journal={IEEE Transactions on Medical Imaging}, 
  title={{RETOUCH: The Retinal OCT Fluid Detection and Segmentation Benchmark and Challenge}}, 
  year={2019},
  volume={38},
  number={8},
  pages={1858-1874},
  doi={10.1109/TMI.2019.2901398}
}

@article{compare1,
  title={Automatic segmentation and classification methods using optical coherence tomography angiography (OCTA): A review and handbook},
  author={Meiburger, Kristen M and Salvi, Massimo and Rotunno, Giulia and Drexler, Wolfgang and Liu, Mengyang},
  journal={Applied Sciences},
  volume={11},
  number={20},
  pages={9734},
  year={2021},
  publisher={MDPI}
}

@article{compare2,
  title={Automatic blood vessels segmentation based on different retinal maps from OCTA scans},
  author={Eladawi, Nabila and Elmogy, Mohammed and Helmy, Omar and Aboelfetouh, Ahmed and Riad, Alaa and Sandhu, Harpal and Schaal, Shlomit and El-Baz, Ayman},
  journal={Computers in biology and medicine},
  volume={89},
  pages={150--161},
  year={2017},
  publisher={Elsevier}
}

\end{document}